\documentclass[11pt,a4paper]{article}
\pdfoutput=1
\usepackage[utf8]{inputenc}
\usepackage{amsmath}
\usepackage{amsfonts}
\usepackage{amssymb}

\usepackage{jinstpub}

\usepackage{physics}
\usepackage{subfigure}
\usepackage{siunitx}

\title{Measurements with a TRISTAN prototype detector system at the "Troitsk nu-mass" experiment in integral and differential mode}

\author[1,2]{Tim~Brunst}
\author[1,2]{Thibaut~Houdy}
\author[1,2]{Susanne~Mertens}
\author[3,4]{Aleksander~Nozik}
\author[3]{Vladislav~Pantuev}
\author[3]{Djohnrid~Abdurashitov}
\author[1]{Konrad~Altenmüller}
\author[3]{Alexander~Belesev}
\author[5]{Luca~Bombelli}
\author[3]{Vasiliy~Chernov}
\author[3]{Evgeniy~Geraskin}
\author[6]{Anton~Huber}
\author[3]{Nikolay~Ionov}
\author[4]{Gregory~Koroteev}
\author[6]{Marc~Korzeczek}
\author[1,7]{Thierry~Lasserre}
\author[8]{Peter~Lechner}
\author[3]{Nikolay~Likhovid}
\author[3,9]{Alexey~Lokhov}
\author[3]{Vladimir~Parfenov}
\author[1,2]{Daniel~Siegmann}
\author[3]{Aino~Skasyrskaya}
\author[2]{Martin~Slezák}
\author[3]{Igor~Tkachev}
\author[3]{Sergey~Zadorozhny}

\affiliation[1]{Technical University of Munich, Arcisstraße 21, 80333 München, Germany}
\affiliation[2]{Max Planck Institute for Physics, Föhringer Ring 6, 80805 München, Germany}
\affiliation[3]{Institute for Nuclear Research of Russian Academy of Sciences, Prospekt 60-letiya Oktyabrya 7a, Moscow 117312, Russian Federation}
\affiliation[4]{Moscow Institute of Physics and Technology, 9 Institutskiy per., Dolgoprudny, Moscow Region, 141700, Russian Federation}
\affiliation[5]{XGLab srl, Bruker Nano Analytics, Via Conte Rosso 23, 20134 Milano, Italy}
\affiliation[6]{Karlsruhe Institute of Technology, Hermann-von-Helmholtz-Platz 1, 76344 Eggenstein-Leopoldshafen, Germany}
\affiliation[7]{Centre CEA de Saclay, 91191 Gif-sur-Yvette cedex, France}
\affiliation[8]{Halbleiterlabor of the Max Planck Society, Otto-Hahn-Ring 6, 81739 München, Germany}
\affiliation[9]{Institute for Nuclear Physics, University of Muenster, Wilhelm-Klemm-Str. 9, 48149 Münster, Germany}

\emailAdd{tim.brunst@mpp.mpg.de}

\abstract{Sterile neutrinos emerge in minimal extensions of the Standard Model which can solve a number of open questions in astroparticle physics. For example, sterile neutrinos in the keV-mass range are viable dark matter candidates. Their existence would lead to a kink-like distortion in the tritium $\upbeta$-decay spectrum. In this work we report about the instrumentation of the Troitsk nu-mass experiment with a 7-pixel TRISTAN prototype detector and measurements in both differential and integral mode. The combination of the two modes is a key requirement for a precise sterile neutrino search, as both methods are prone to largely different systematic uncertainties. Thanks to the excellent performance of the TRISTAN detector at high rates, a sterile neutrino search up to masses of about 6~keV could be performed, which enlarges the previous accessible mass range by a factor of 3. Upper limits on the neutrino mixing amplitude in the mass range $<$~5.6~keV (differential) and $<$~6.6~keV (integral) are presented. These results demonstrate the feasibility of a sterile neutrino search as planned in the upgrade of the KATRIN experiment with the final TRISTAN detector and read-out system.}


\arxivnumber{1909.02898}

\notoc

\begin{document}

\maketitle
\flushbottom

\section{Introduction}

Despite strong evidences of the existence of Dark Matter (DM) in our universe~\cite{Rubin1980} a direct detection is still pending. Several theories beyond the Standard Model (SM) consider the existence of a right-handed partner of the left-handed neutrino in order to generate neutrino masses. This right-handed neutrino would be sterile regarding the weak interaction. Many mass ranges for the corresponding "sterile" mass eigenstate are considered~\cite{Adhikari2017}, mostly from eV to GeV.\footnote{The existence of a sterile neutrino flavour eigenstate would imply an additional fourth mass eigenstate. Throughout this paper, "the mass of the sterile neutrino" refers to this new mass eigenstate.} The keV-scale sterile neutrino is a viable candidate for cold and warm DM~\cite{Merle2012}.\\

The existence of such a particle would have dramatic consequences on astrophysical observations. Telescope experiments, that search for a mono-energetic x-ray line arising from the decay of relic sterile neutrinos, constrain the active-sterile mixing amplitude to $\sin^2(\theta)<10^{-6}$--$10^{-8}$~\cite{Watson2012,Boyarsky2008}. Stringent limits are also set by cosmological considerations and observations of structure formation in the early universe~\cite{Boyarsky2012}. Finally, laboratory-based experiments provide a direct search for their existence~\cite{Hiddemann1995, Holzschuh1999, Abdurashitov2017}.\\

A widely used method for a direct detection of sterile neutrinos is based on high-precision measurements of the energy spectrum of $\upbeta$-decay or electron capture. Experiments like KATRIN~\cite{Angrik2005}, Project-8~\cite{Esfahani2017}, HOLMES~\cite{Faverzani2016} or ECHo~\cite{Gastaldo2017} aim at determining the effective electron (anti-)neutrino mass by measuring the spectral shape close to the endpoint, where the neutrino mass effect on the spectral shape is maximal. In a $\upbeta$-decay a $\bar\upnu_\mathrm{e}$ and a $\upbeta$-electron are emitted. The released energy is shared between the daughter molecule, the electron, and the electron anti-neutrino. The current best limit from a direct measurement on the effective electron anti-neutrino mass was reached with this method~\cite{Tanabashi2018}.\\

\begin{figure}
    \centering
    \includegraphics[width=0.6\textwidth]{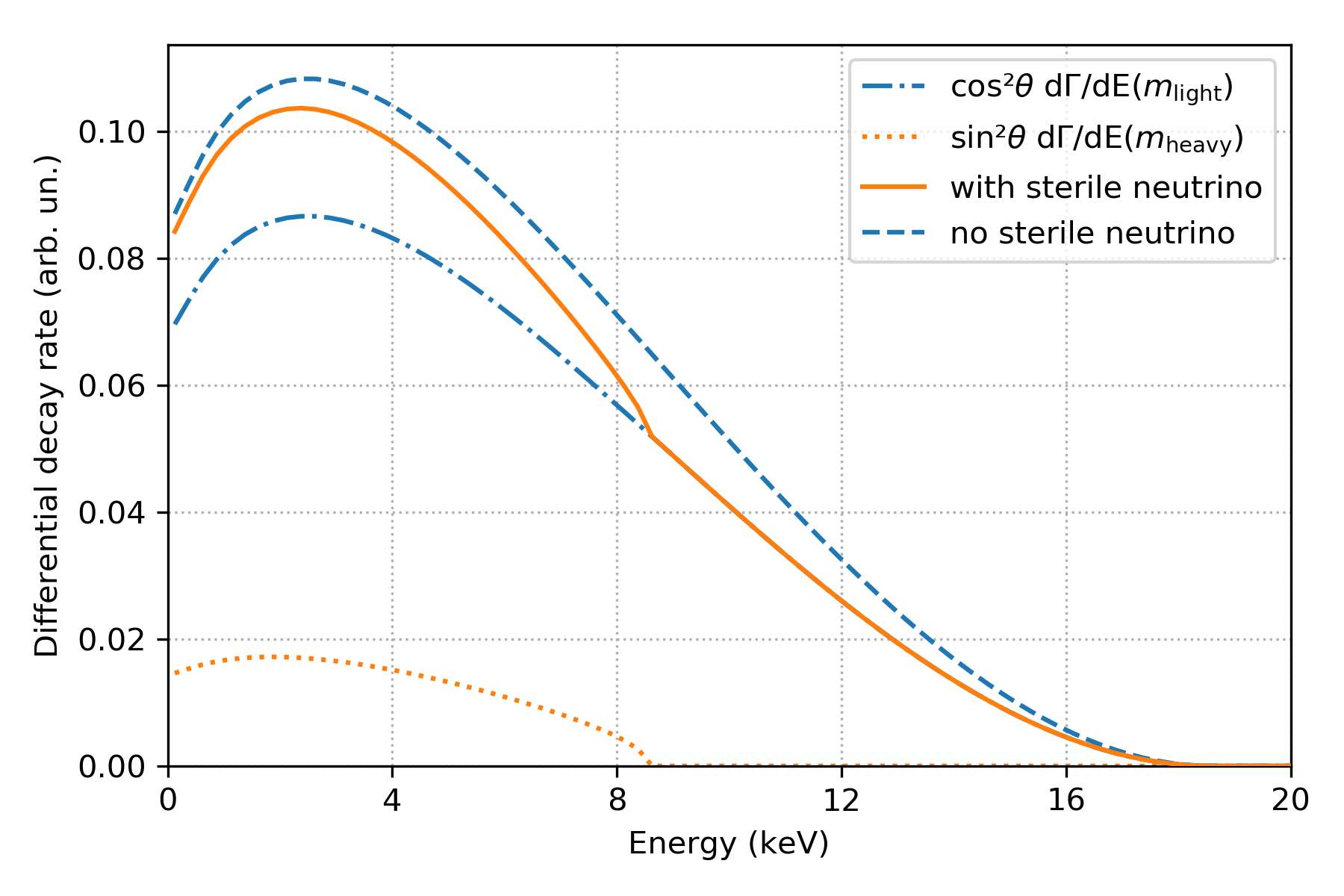}
    \caption{\textbf{Tritium $\upbeta$-decay spectrum with sterile neutrino admixture.} The spectrum is dominated by the known light neutrino mass states $m_\mathrm{light}$. The admixture of a sterile neutrino is generated with a mass of $m_\mathrm{heavy}=10$~keV and a mixing amplitude of $\sin^2(\theta)=0.2$ (exaggerated for illustration).}
    \label{fig:beta_spectrum}
\end{figure}

The signature of a keV-scale sterile neutrino would manifest itself as a kink and a broad distortion of the resulting continuous electron energy spectrum~\cite{Adhikari2017}
\begin{equation}
    \dv{\Gamma}{E} = \cos^2(\theta)\dv{\Gamma(m_\mathrm{light})}{E} + \sin^2(\theta)\dv{\Gamma(m_\mathrm{heavy})}{E}~.
    \label{equ:tritium_spectrum}
\end{equation}
Figure~\ref{fig:beta_spectrum} illustrates this effect on the $\upbeta$-spectrum when choosing tritium as electron source. While the position of the kink depends on the mass of the sterile neutrino $m_\mathrm{heavy}$, the signal strength of the broad distortion is governed by the active-sterile mixing amplitude $\sin^2(\theta)$. Tritium is well suited because of its low endpoint energy of 18.6~keV and its comparatively short half-life of 12.3~years, enabling to produce a high-luminosity radioactive source.\\

One promising proposal to search for a keV-scale sterile neutrino signature is to operate the KATRIN experiment extended by the TRISTAN detector~\cite{Mertens2015}. As the KATRIN neutrino mass program will continue until at least 2023, the Troitsk nu-mass experiment is the only KATRIN-like installation in operation that is capable for a keV-scale sterile neutrino search using tritium $\upbeta$-decay.

\section{Troitsk nu-mass experiment}\label{sec:nu_mass}

\begin{figure}
    \centering
    \includegraphics[width=0.9\textwidth]{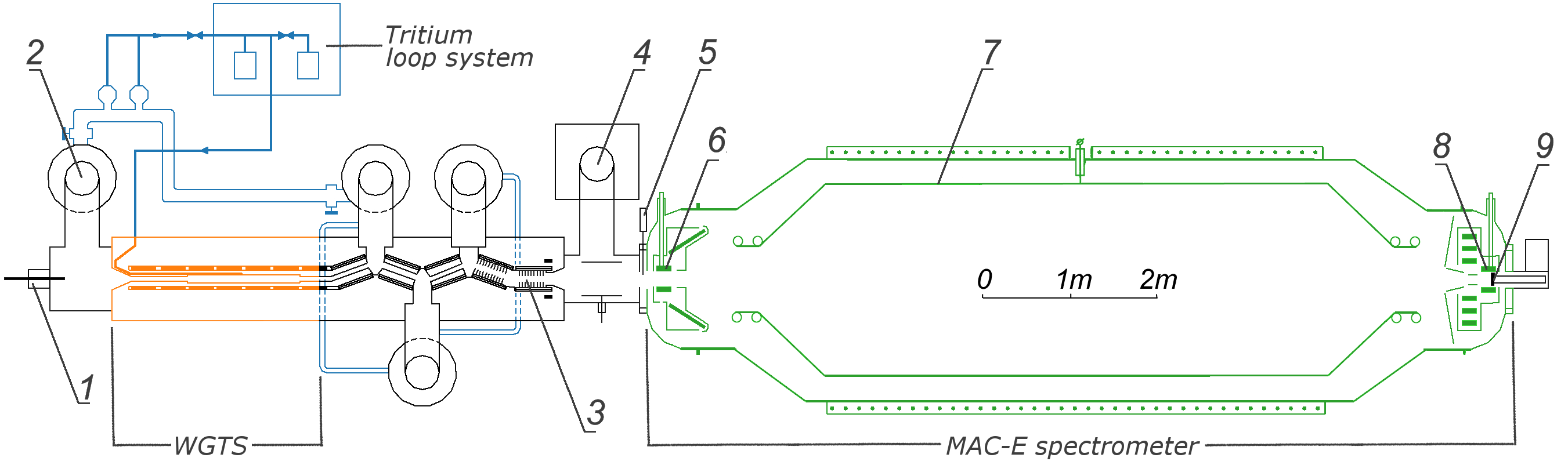}
    \caption{\textbf{Scheme of the Troitsk nu-mass experiment.}
    blue: tritium loop system; orange: WGTS; green: MAC-E spectrometer; 1: electron gun; 2: diffusion pumps; 3: argon cryo-pump; 4: Ti-pump; 5: fast shutter; 6: pinch magnet; 7: spectrometer electrode; 8: detector magnet; 9: detector. Adapted from~\cite{Abdurashitov2015}.}
    \label{fig:troitsk_scheme}
\end{figure}

The Troitsk nu-mass experiment is conducted by the Institute for Nuclear Research of the Russian Academy of Sciences. Besides the Mainz experiment, it is one of the two technological predecessors that joined the KATRIN collaboration. Both of them still hold the best limits of $m_\nu < 2.05$~eV~\cite{Aseev2011} and $2.3$~eV~\cite{Kraus2005} (95~\% C.L.) via a direct measurement of the neutrino mass. The setup consists of two main components: the Windowless Gaseous Tritium Source (WGTS) and the Magnetic Adiabatic Collimation spectrometer combined with an Electrostatic filter (MAC-E). A scheme of the entire setup is displayed in figure~\ref{fig:troitsk_scheme}. The source activity is lower and the spectrometer resolution is worse than in the KATRIN design but the experimental principle is the same. Tritium is injected in form of deuterium-tritium molecules (DT) in the middle of the WGTS and pumped out at both sides. To avoid a potential source of background, no tritium must reach the main spectrometer. Electrons from the decay are guided magnetically towards the spectrometer and detector section. Magnets at both ends of the WGTS create a magnetic bottle in which electrons starting with a polar angle larger than $24^\circ$ are trapped and consequently prevented from entering the spectrometer section.\\

A retarding potential $U$ in the center of the spectrometer acts as a high-pass filter for electrons from the tritium decay. However, it is only sensitive to the longitudinal momentum component. The magnetic field inside the MAC-E spectrometer is strong at both ends (7.2 and 3~T) and weak (3~mT) in the center. As the electron's magnetic moment is conserved in an adiabatic motion, its transversal momentum component is transformed into the longitudinal one when travelling from the strong into the weak field. This strongly improves the resolution of the spectrometer~\cite{Lobashev1985}, e.g.~at $E=18$~keV to
\begin{equation}
    \Delta E_{\bot}=E\cdot\frac{B_\mathrm{min}}{B_\mathrm{max}}\approx1.5~\mathrm{eV}~.
    \label{equ:spec_res}
\end{equation}
All electrons that overcome the retarding energy $qU$ are counted by the detector. By varying the retarding energy, an integral spectrum is measured.\\

Since the neutrino mass program that ended in 2004, Troitsk nu-mass has been upgraded in order to search for the signature of a keV-scale sterile neutrino~\cite{Abdurashitov2015}. The latest results set new upper limits to the active-sterile mixing amplitude $\sin^2(\theta)$ in the mass range of 0.1 to 2~keV~\cite{Abdurashitov2017}. Events under detection threshold and electronics dead time were identified as dominant sources of systematic uncertainties, both arising from the detector system. By using the novel detector system of TRISTAN, Troitsk nu-mass overcomes its dominant source of systematic uncertainty on the one hand and on the other hand a sterile neutrino search also on the differential spectrum becomes possible.

\section{TRISTAN detector}\label{sec:tristan}

The TRISTAN group is part of the KATRIN collaboration. KATRIN is designed to probe the absolute neutrino mass scale in a direct laboratory-based measurement \cite{Angrik2005}. Since the neutrino mass manifests itself only at the endpoint region of the $\upbeta$-decay, KATRIN scans the retarding energy down to only about 100~eV below the tritium endpoint at 18.6~keV where the signal rate at the detector is low ($<$~100~kcps). The goal of the TRISTAN project is to utilize the KATRIN experiment to search for the signature of a keV-scale sterile neutrino. This kink-like signature may appear anywhere in the spectrum, depending on the mass of the sterile neutrino (see figure~\ref{fig:beta_spectrum}). Therefore, the TRISTAN project aims at extending the measurement interval down to several keV below the endpoint. In this case, the count rate is increased by several orders of magnitude compared to KATRIN standard operation. Since the KATRIN detector is not designed to handle such high count rates, the TRISTAN group is currently developing a novel detector system.\\

In order to be sensitive to the possible tiny ($<$~ppm-level) sterile neutrino signal, the requirements on the TRISTAN detector system are a good energy resolution ($<$~300~eV (fwhm) at 20~keV), a low energy detection threshold (1~keV) and the ability to manage high count rates ($\mathcal{O}(10^8)$~cps). Therefore, the detector chip will be segmented into an array of more than 3000 pixels to reduce the count rate per pixel below 100~kcps. To determine the ideal pixel diameter, detailed Monte Carlo (MC) simulations including particle tracking were performed using the simulation software \textit{Kassiopeia}~\cite{Groh2015} and \textit{KESS}~\cite{Renschler2011}. Taking into account the exact geometry of the KATRIN vacuum system, electron backscattering from the detector, back-reflection at electromagnetic fields, and interactions inside the silicon detector, the pixel diameter was optimized to 3~mm. This results in a total number of 3486 pixels to cover the whole electron beam tube~\cite{Mertens2019}. Silicon drift detectors (SDD) are optimized for high count rates and large-area coverage~\cite{Gatti1984}. Despite large pixels, the small anode capacitance maintains a high energy resolution. Thus, the SDD is the ideally suited technology for the TRISTAN detector.\\

\begin{figure}
    \centering
    \includegraphics[width=0.45\textwidth]{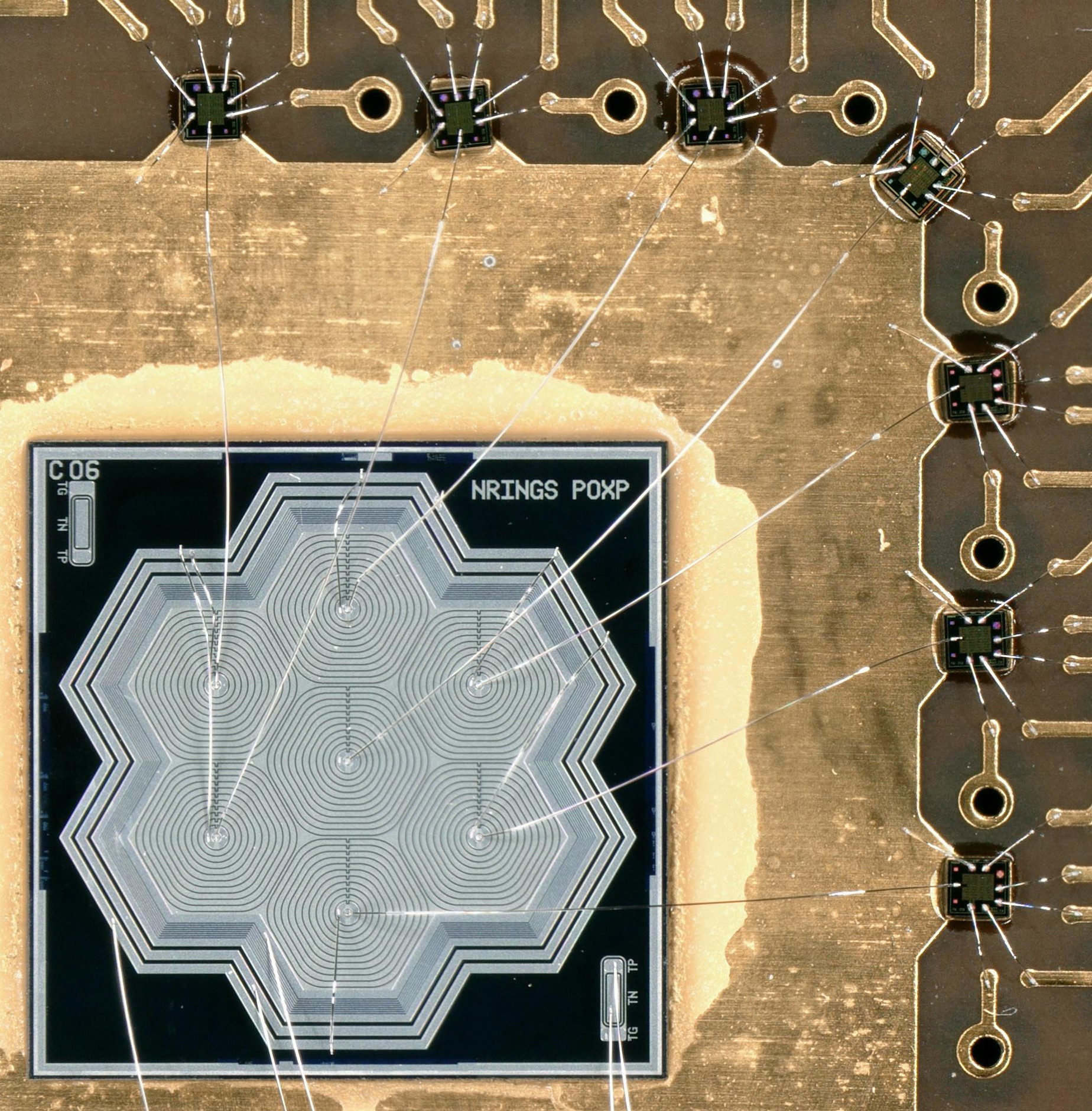}
    \caption{\textbf{Close-up photograph of a TRISTAN detector chip.} The photograph shows the read-out side of the detector 2~mm pixel diameter chip. Also shown are  the seven CUBE pre-amplifier ASICs to which the anodes of the pixels are bonded to.}
    \label{fig:detector}
\end{figure}

Several SDD chips with seven hexagonal pixels in different diameters (0.25, 0.5, 1 and 2~mm) were produced at the semiconductor laboratory of the Max Planck Society (HLL)~\cite{Lechner2001}. One of them is shown in figure~\ref{fig:detector}. Pre-amplifiers and data acquisition system (DAQ) options are currently under investigation. A detailed characterization campaign was performed using a read-out system provided by XGLab.\footnote{www.xglab.it} The front-end electronics consist of a CUBE~\cite{Bombelli2011} pre-amplifier application-specific integrated circuit (ASIC) for each pixel. It is a charge-sensitive pre-amplifier, operated in pulsed-reset mode. The 8-channel DANTE digital pulse processing system is used as back-end electronics. The full signal waveform is digitized and the energy is estimated through a typical trapezoidal filter. Each channel is recorded with a sampling rate of 125~MHz and 16-bit resolution. All channels share an internal clock for time synchronization. An excellent performance of the prototype system was demonstrated in the laboratory~\cite{Mertens2019}. In this work we demonstrate the feasibility of operating the system in realistic conditions.\\

\section{Measurements with TRISTAN at Troitsk nu-mass}

The TRISTAN detector prototype system used for the measurements presented in this work consists of a 2~mm pixel diameter chip equipped with the XGLab read-out system, described in section~\ref{sec:tristan}. The Troitsk nu-mass standard detector was removed from the detector section and exchanged by the TRISTAN system. Thanks to its excellent performance at high rates, it was possible to operate the Troitsk nu-mass experiment in both an integral and for the first time a differential mode:
\begin{itemize}
    \item \textbf{Integral mode:}
    The detector counts the electrons that pass through the spectrometer depending on the retarding energy setting. By scanning the retarding energy over a voltage range an integral tritium spectrum is measured. The energy resolution is determined by the spectrometer and the detector only counts incoming events. This is the standard mode of the Troitsk nu-mass experiment.
    
    \item \textbf{Differential mode:}
    The retarding energy is set to a low value so that electrons of a wide energy range pass the spectrometer. Their energy is analyzed by the detector and a differential tritium spectrum is measured. In this case, the energy resolution is determined by the detector. Given the energy resolution of the standard Troitsk nu-mass detector of around 3~keV, it was not possible to analyze its differential spectrum for a signature of a keV-scale sterile neutrino. The TRISTAN detector, in contrast, provides a resolution at the 300~eV level.
\end{itemize}

\subsection{Measurement in integral mode}\label{sec:meas_int}

The retarding energy was scanned between 12~keV and 18.6~keV in 50~eV steps. At each of those \textit{points}, a measurement of the differential spectrum was taken for 30~s. The count rate varied between 5.3~kcps and 0.1~cps depending on the retarding energy. In order to monitor the tritium activity, each $8^\mathrm{th}$ point was taken at a retarding energy of 15~keV. The scan includes 153 points at different retarding energies, 21 of which are monitor points. The duration of the scan was 76.5~min live time. In total 5.8~million electrons were detected. Additionally, measurements with the electron gun were performed to determine the column density of the WGTS. This is an important input for the simulations determining the response of the source section (see section~\ref{sec:response}), as a higher column density increases the probability of scattering and thus of an energy loss of electrons in the source section. Subsequently, a background run was performed without tritium in the WGTS.

\subsection{Measurement in differential mode}\label{sec:meas_diff}

\begin{figure}
    \subfigure[Measured differential tritium spectrum.]{
        \includegraphics[width=0.45\textwidth]{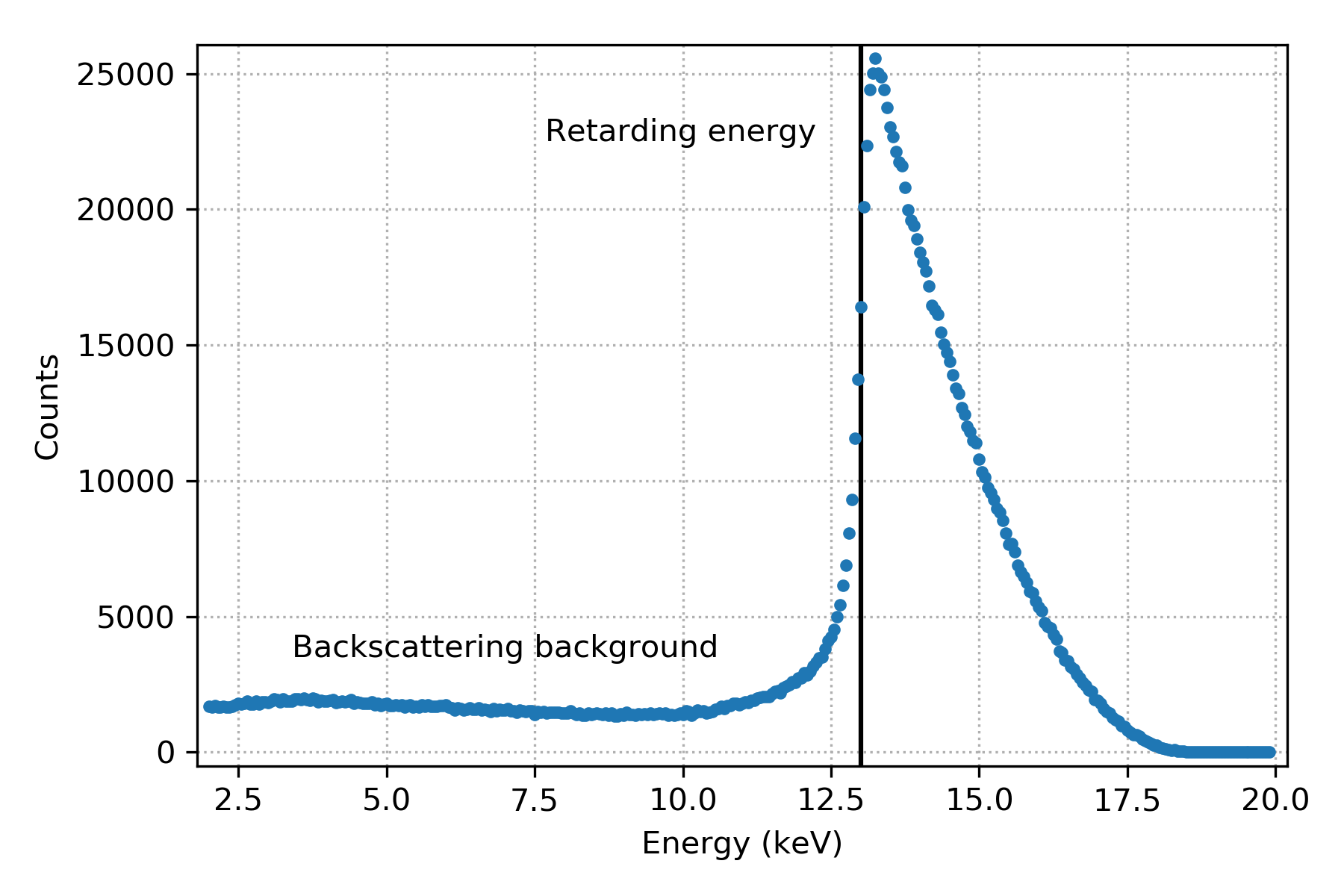}
        \label{fig:Troitsk2_tritiumspec}
    }
    \subfigure[Mono-energetic spectrometer electrode spectra.]{
        \includegraphics[width=0.45\textwidth]{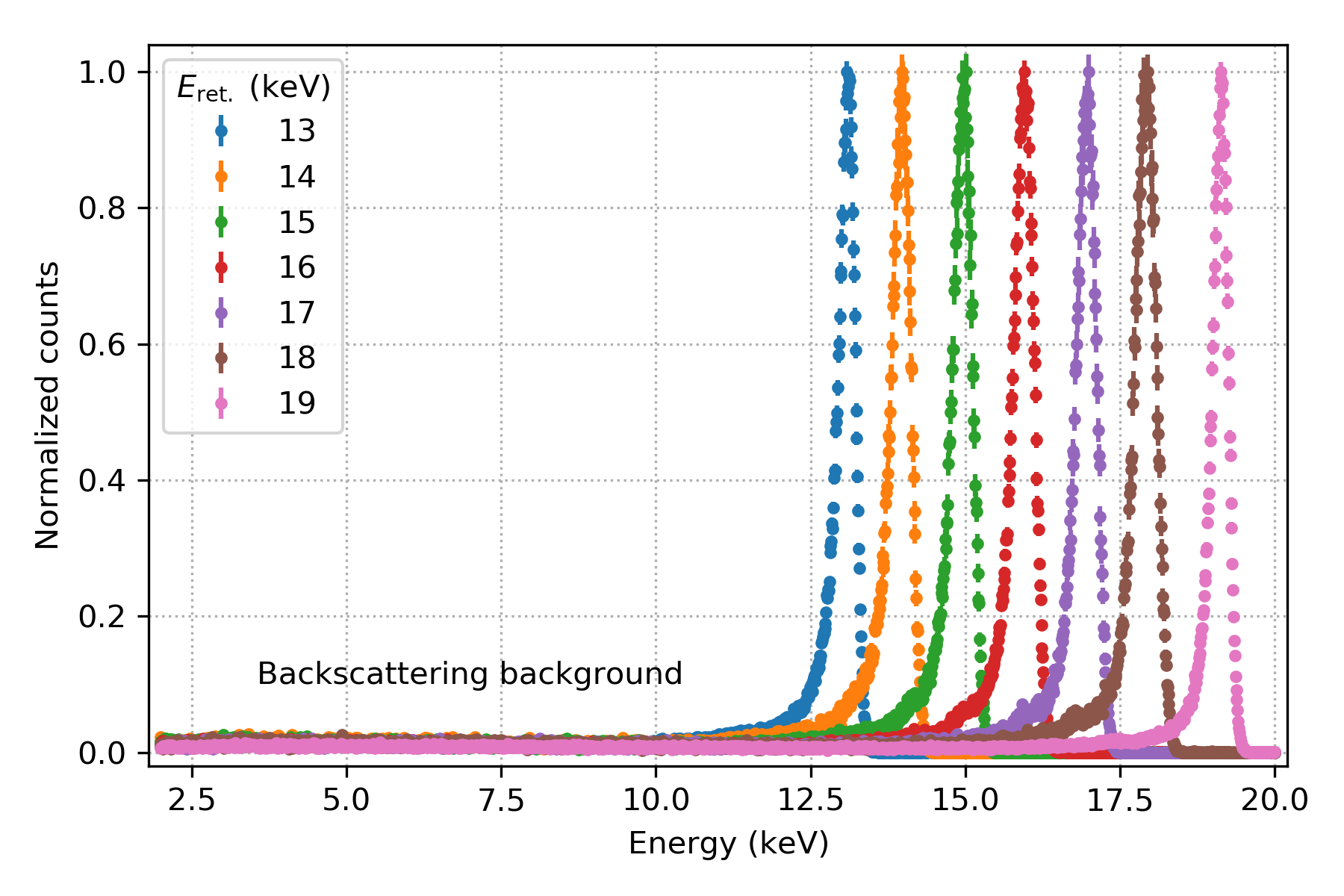}
        \label{fig:Troitsk2_electrodespec}
    }
    \caption{\textbf{Measurements in differential mode.} \protect\subref{fig:Troitsk2_tritiumspec} Indicated is the retarding energy of 13~keV that cuts off the low energy part of the tritium spectrum. Events below this energy are summarized as the backscattering background. They mainly result from partial energy deposition through backscattering but also from energy loss in the detector dead layer and undetected charge sharing between pixels. \protect\subref{fig:Troitsk2_electrodespec} The spectra are normalized to their respective maximum. The backscattering background can also be seen here.}
    \label{fig:Troitsk2_measurement}
\end{figure}

The retarding energy was fixed to 13~keV and the energy of each event from tritium decay was measured by the detector over about 5~h in total. The count rate was around 100~cps. A total of 1.7~million electrons was detected. A crucial input for the analysis is the detector response function (see section~\ref{sec:response}). To this end, measurements of mono-energetic electrons emitted by the spectrometer electrode were carried out. In this configuration, the fast shutter is closed and the pinch magnet is turned off (see figure~\ref{fig:troitsk_scheme}). Electrons are emitted by the spectrometer electrode via impact ionization of residual ions and guided by the magnetic field of the detector magnet onto the detector chip. These electrons carry exactly the energy corresponding to the electrode voltage with a sub-eV width when they reach the detector. Mono-energetic responses were recorded from 13~keV to 19~keV in 1~keV steps. The energy resolution for electrons was $\Delta E =$ 236--363~eV (fwhm) with a peaking time of $\SI{0.8}{\micro\second}$. The measured differential tritium spectrum and all measured mono-energetic spectra from the spectrometer electrode are shown in figure~\ref{fig:Troitsk2_measurement}. The electron gun does not allow for a homogeneous illumination of the detector and thus cannot be used to determine the response.

\section{Data analysis}

In this chapter, we first describe the inputs which are common to both the integral and differential analysis: the underlying $\upbeta$-decay model and the experimental response. Thereafter, we focus on the individual fits of integral and differential spectra.

\subsection{\boldmath$\upbeta$-decay model}\label{sec:model}

The model for the differential tritium spectrum
\begin{equation}
    \dv{\Gamma(m_\upnu)}{E} \propto S(E, E_0)\cdot F(Z,E)\cdot (E+m_\mathrm{e})\cdot p\cdot \sum_i P_i\cdot (E_0-E-E_i)^2\cdot \sqrt{1-\biggl(\frac{m_\upnu}{E_0-E-E_i}\biggr)^2}
    \label{equ:fermi_theory}
\end{equation}
is composed of the Fermi function $F$~\cite{Simpson1981}, small term corrections $S$~\cite{Wilkinson1991} and an energy-dependent final state distribution with probabilities $P_i$ and energy states $E_i$~\cite{Schaetz2019}. Furthermore, there is the kinetic energy $E$, momentum $p$, and mass $m_\mathrm{e}$ of the electron, the effective electron neutrino mass $m_\upnu$ and the spectral endpoint energy $E_0$, which is the maximal electron kinetic energy in case $m_\upnu=0$.\\

A sterile neutrino is considered in the following way. If the electron neutrino flavor contains an additional heavy (mostly sterile) mass eigenstate, the total electron spectrum is given as a superposition of spectra with different neutrino mass components. The dominant component corresponds to the active neutrino, whose mass is neglected here, and a much smaller component, corresponding to the heavier sterile neutrino mass. This introduces only two additional parameters: active-sterile mixing amplitude $\sin^2(\theta)$ and sterile mass $m_\mathrm{heavy}$, as shown in equation~\ref{equ:tritium_spectrum}.

\subsection{Experimental response}\label{sec:response}

In the most general case, a response gives the probability for an electron to change its energy, direction of motion, and/or position. The response of the entire Troitsk nu-mass setup is separated into three parts --- trapping, transport, and detector --- and expressed in response matrices. These response matrices describe the energy change of mono-energetic electrons in the respective part of the experimental setup for all possible energies. The matrices are convolved with the model in a matrix multiplication prior to the least-squares fit of the data.

\paragraph{Trapping}
The magnetic field inside the WGTS is 0.6~T and rises up to 3.7~T at both of its ends. This field configuration creates a magnetic trap which confines electrons with large polar angles to the WGTS. Some electrons escape this trap and reach the spectrometer after inelastic scattering and a large angular change. The scattering leads to a redistribution of electron energies to lower values. This redistribution was determined via MC simulations~\cite{Aseev2011}. Here, electrons scatter at most once during one traverse through the WGTS, which is given by the low column densities in the Troitsk nu-mass source section. The number of trapped electrons per unit energy range that eventually leave the WGTS is constant over the electron's energy loss. The magnitude of this constant depends on the ratio of inelastic to elastic scattering cross-section and thus on the initial electron energy. For each initial energy, the redistribution of energies is simulated and combined in the trapping response matrix $T$.

\paragraph{Transport}
The response of the source and spectrometer components is simulated using a C\texttt{++} based code specially developed for keV sterile neutrino searches. The simulation accounts for the energy loss through scattering in the WGTS and the MAC-E filter transmission considering angular distribution and position of the electrons~\cite{Lokhov2018}. The resulting energies are binned in equidistant steps which yields the transport response matrix $R$.

\begin{figure}
    \subfigure[Parameterization of an electrode measurement.]{
        \includegraphics[width=0.45\textwidth]{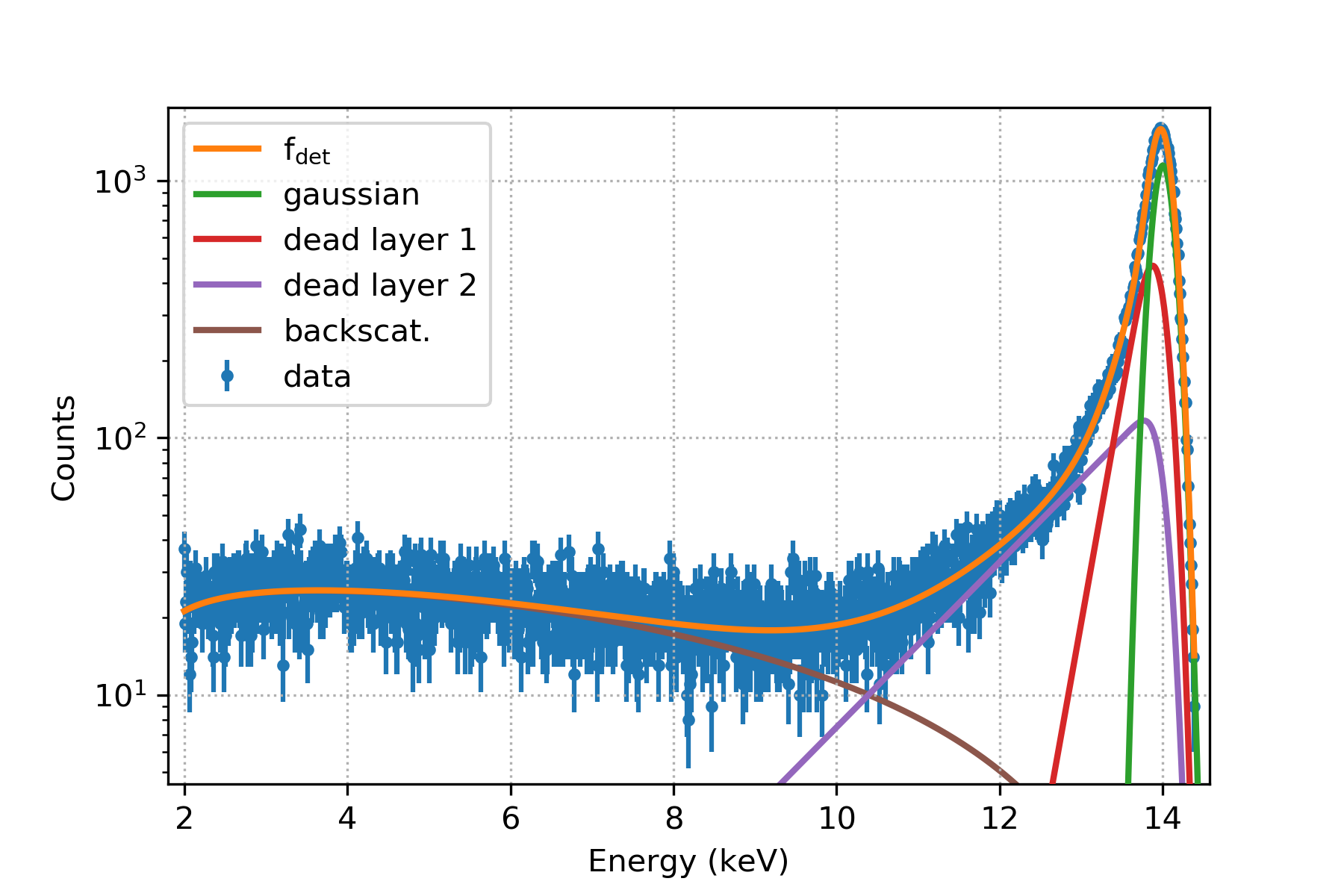}
        \label{fig:parameterization}
    }
    \subfigure[Response matrix for all initial energies.]{
        \includegraphics[width=0.45\textwidth]{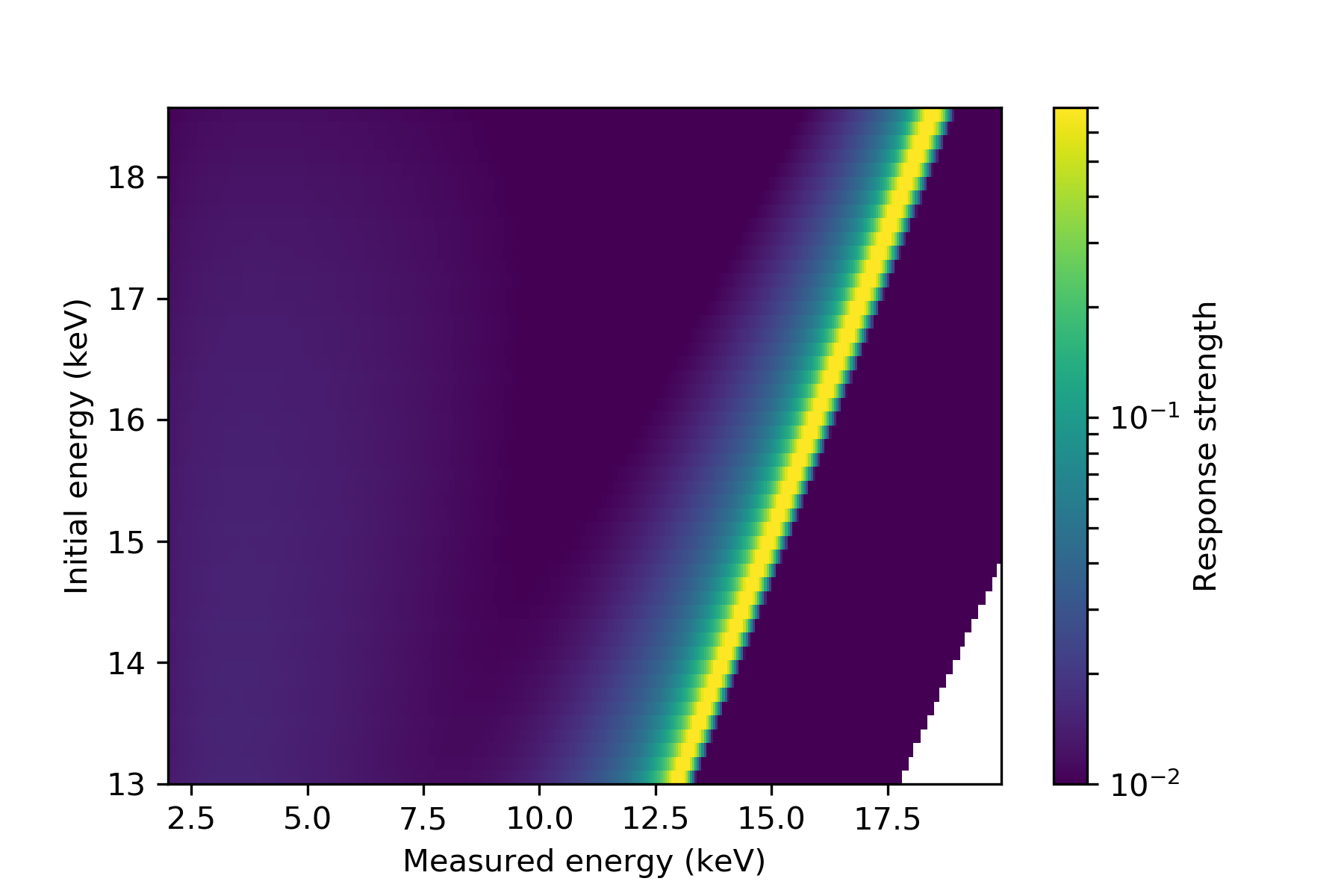}
        \label{fig:response_matrix}
    }
    \caption{\textbf{Detector response obtained from electrode measurements.} The data shown in figure~\protect\subref{fig:parameterization} was taken at a retarding energy of 14~keV. The Gaussian function in equation~\ref{equ:parameterization1} is fit to the main peak of the electron spectrum. The low energy shoulder of the main peak, which arises mainly from the dead layer effect, is described by equation~\ref{equ:parameterization2}. The backscattering background is described by the function in equation~\ref{equ:parameterization3}. Note the logarithmic vertical axis. Figure~\protect\subref{fig:response_matrix} shows the response to all initial energies between 13 and 18.6~keV in form of a matrix. A slice through the matrix at 14~keV initial energy yields the plot in figure~\protect\subref{fig:parameterization}. }
    \label{fig:initial_responses}
\end{figure}

\paragraph{Detector} 
The detector response is used in the analysis of the differential mode only. It is determined with the help of the measured responses to mono-energetic electrons from the spectrometer electrode. Those measured spectra are parameterized and fit with the empirical function~\cite{Altenmueller2019}
\begin{align}
    \label{equ:parameterization1}
    f_\mathrm{det}(E; \mu, \sigma, n_0, n_1, n_2, n_3, \beta_1, \beta_2, a, b) = 
    \underbrace{n_0\cdot\operatorname{e}^{-\bigl(\frac{\mu-E}{\sqrt{2}\sigma}\bigr)^2}}_\mathrm{Gaussian} + \\
    \label{equ:parameterization2}
    \sum_{i=1}^{2}\underbrace{n_0\cdot n_i\cdot\operatorname{e}^{-\frac{\mu-E}{\beta_i}}\cdot\Biggl(1-\mathrm{erf}\left[-\frac{\mu-E}{\sqrt{2}\sigma}+\frac{\sigma}{\sqrt{2}\beta_i}\right]\Biggr)}_\mathrm{Dead~layer} + \\
    \label{equ:parameterization3}
    \underbrace{\Re\left[n_0\cdot n_3\cdot\biggl(\frac{E-1500}{\mu}\biggr)^a\cdot\biggl(1-\frac{E}{\mu}\biggr)^b\right]\cdot\Theta(\mu-E)}_\mathrm{Backscattering}~,
\end{align}
where $\mu$ is the initial electron energy that equals the retarding energy. This function was specifically developed for this purpose and is able to adjust to spectral changes of the response for various incident energies. Like this, the response can be interpolated and predicted for energies that have not been measured, which is important when dealing with a continuous spectrum like the one from tritium decay. The real part in equation~\ref{equ:parameterization3} prevents this term from getting imaginary during the fit and acts as a detection threshold effect. The function is visualized in figure~\ref{fig:parameterization} and has 10 free parameters. A linear dependence of all parameters on $\mu$ is introduced by fitting a linear function to all obtained function parameter values, e.g.
\begin{equation}
    \beta_2(\mu)=m_{\beta_2}\cdot\mu+c_{\beta_2}
    \label{equ:lin_param_energ_dep}
\end{equation}
with slope $m_{\beta_2}$ and offset $c_{\beta_2}$. An exception is made for two parameters:
\begin{itemize}
    \item[$n_0$] is the normalization and cannot be inferred from electrode measurements. Instead, a simulation was performed to determine the amount of backscattered electrons that are not back-reflected and get lost in the WGTS depending on their initial energy~\cite{Grigorieva2016}.
    \item[$n_3$] does not show a slope and is described by a single constant $c_{n_3}$.
\end{itemize} 
This yields a total of 15 free parameters $m_{\sigma}$, $c_{\sigma}$, $m_{n_1}$, $c_{n_1}$, $m_{n_2}$, $c_{n_2}$, $m_{\beta_1}$, $c_{\beta_1}$, $m_{\beta_2}$, $c_{\beta_2}$, $m_{a}$, $c_{a}$, $m_{b}$, $c_{b}$, and $c_{n_3}$, which are in the following referred to as $\phi$. The energy dependence is used to calculate the response for all initial energies that are combined in the detector response matrix $D$, displayed in figure~\ref{fig:response_matrix}.

\subsection{Analysis of the integral mode}

\begin{figure}
    \subfigure[Subtraction of the structure.]{
        \includegraphics[width=0.45\textwidth]{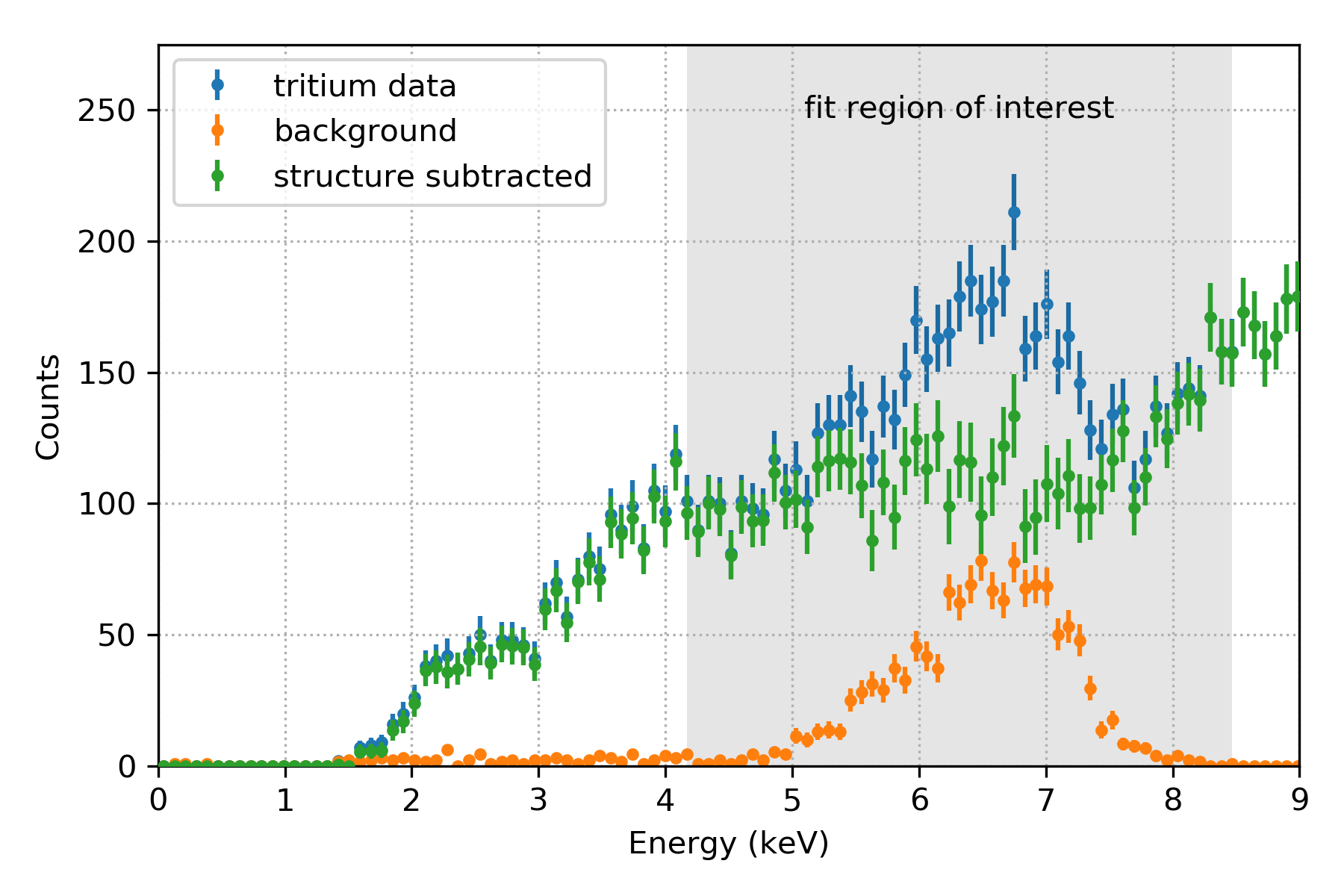}
        \label{fig:bump_subtracted}
    }
    \subfigure[Correction of events below threshold.]{
        \includegraphics[width=0.45\textwidth]{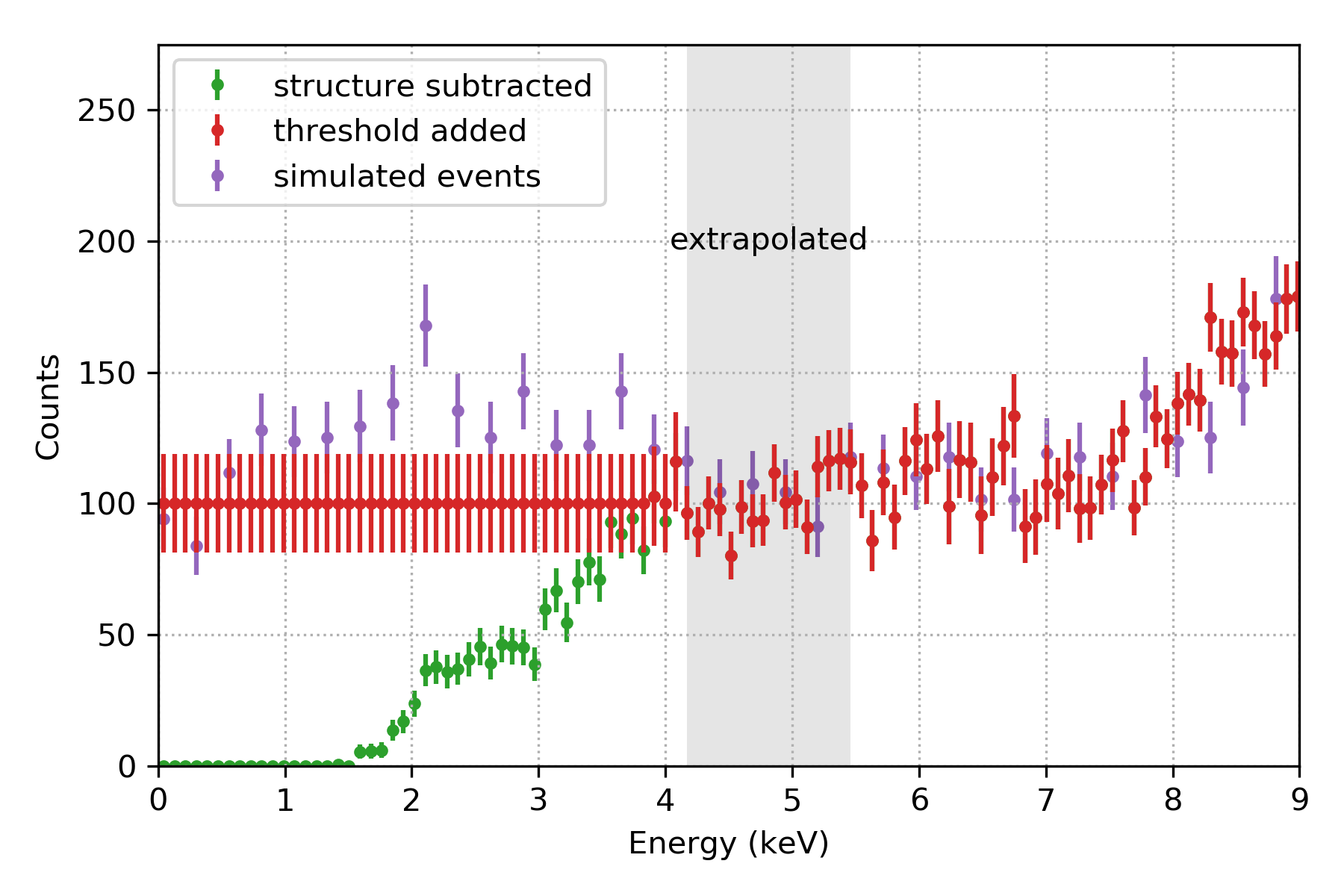}
        \label{fig:threshold_added}
    }
    \caption{\textbf{Corrections on the differential spectrum.} \protect\subref{fig:bump_subtracted} The region in which the background (orange) is fit to the data is shadowed. However, as the peak-like structure has a tail to lower energies the whole background spectrum is subtracted from the tritium data (blue). The resulting spectrum is shown in green. \protect\subref{fig:threshold_added} The region of data including subtraction of the structure (green) which is extrapolated from is shadowed. The resulting spectrum including events below detection threshold is shown in red. The purple spectrum is a simulation of $10^5$ events using Kassiopeia.}
    \label{fig:corrected_spectrum_diff}
\end{figure}

In the case of an integral measurement, the count rate as a function of retarding energy is recorded. The TRISTAN detector, however, enables to measure the differential shape of the tritium spectrum with a resolution high enough to account for features that have not been visible with the standard Troitsk nu-mass detector. Hence, additional corrections on the differential spectrum are conducted before it is integrated to obtain the total count rate. The additional source of uncertainty due to the parameterization of the differential spectrum is avoided and the respective corrections are conducted on the data directly.
\begin{itemize}    
    \item \textbf{Events below threshold:}
    Events below the detection threshold are added using a linear extrapolation of the backscattering tail towards zero energy, depicted in figure~\ref{fig:threshold_added}. The spectral shape of these events was validated by MC simulations using the software package \textit{Kassiopeia}~\cite{Furse2017}. After correcting these effects, all counts in a differential spectrum are summed up to obtain the data point of one retarding energy setting in the integral spectrum.
    
    \item \textbf{Low energy structure:}
    A peak-like structure at retarding energies below 16~keV was seen below the respective retarding energy. These events supposedly originate from a Penning trap in the down-stream section of the spectrometer and are subtracted from the tritium data, as shown in figure~\ref{fig:bump_subtracted}. The shape of the structure is extracted from background measurements while position and amplitude are fit in each point.
    
    \item \textbf{Rate corrections:}
    A rate-dependent dead time correction is applied to account for read-out electronics and vetoed events. The DAQ dead time is given by the peaking and gap time of the trapezoidal energy filter and the recovery time after each reset. Additional dead time due to the multiplicity cut is induced. Furthermore, the tritium loss over the course of the acquisition must be taken into account. To do so, the monitor points are fit linearly. From the obtained function, correction factors for each measurement point are derived.
\end{itemize}

\subsubsection{Systematic uncertainties}\label{sec:integral_systematics}

\begin{figure}
    \centering
    \includegraphics[width=0.6\textwidth]{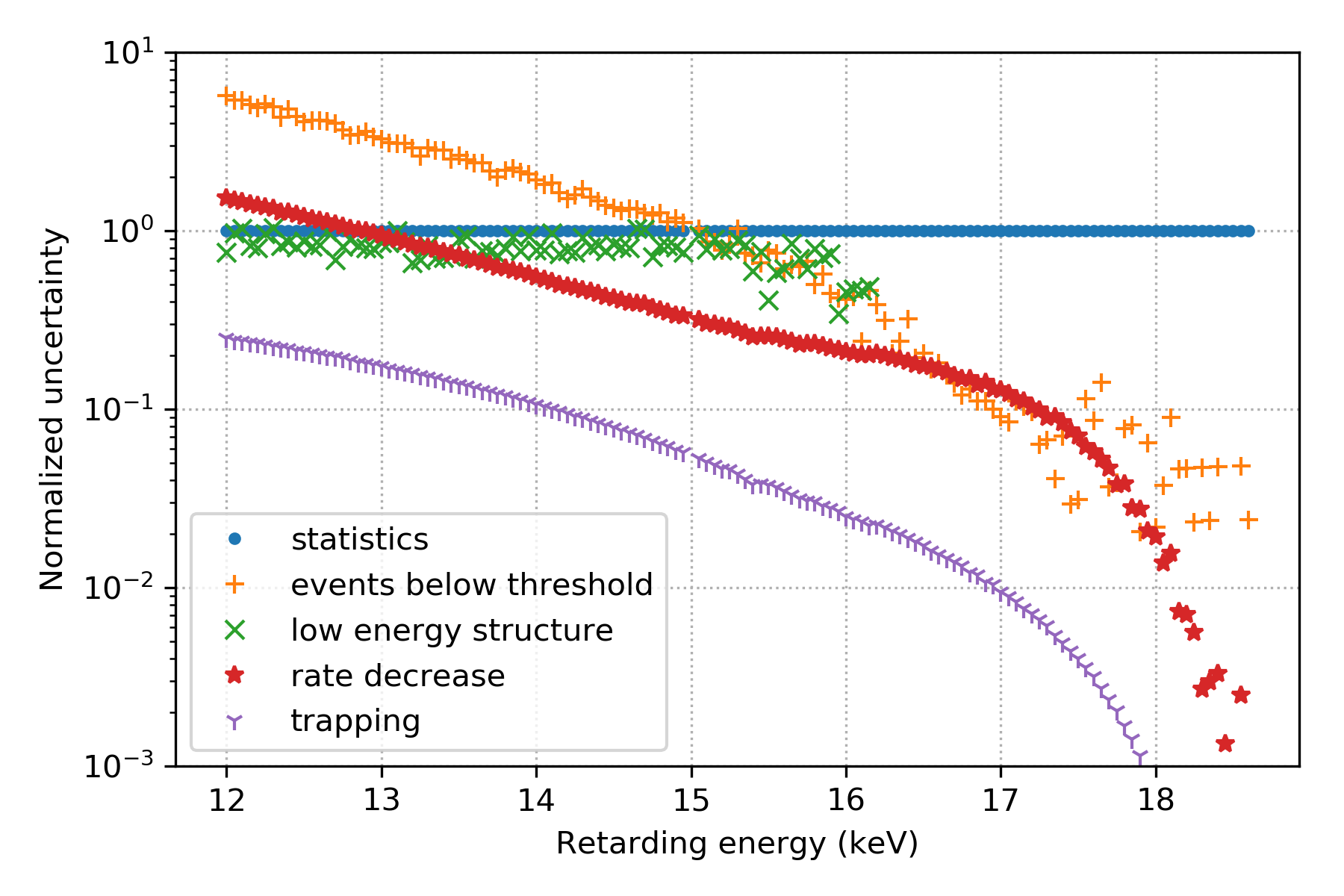}
    \caption{\textbf{Error budget.} Shown is the ratio of uncorrelated systematic to statistical uncertainties. Systematic uncertainties, mostly induced by the events below threshold correction, dominate below 15~keV.}
    \label{fig:error_budget}
\end{figure}

For the most dominant uncertainties of the integral mode, covariance matrices were generated taking point-to-point correlated uncertainties into account. For this purpose, the sample covariance of $10^4$ tritium model spectra was created varying the respective correction parameters within their correlated uncertainties. The uncorrelated part of the uncertainties for all relevant systematic effects are displayed in figure~\ref{fig:error_budget}.
\begin{itemize}
    \item \textbf{Events below threshold:}
    Deviations from a linear extrapolation are considered by adding a quadratic term to the correction function. Its coefficient was varied within a range that covers the deviation obtained from the MC simulation, shown in figure~\ref{fig:threshold_added}.
    
    \item \textbf{Low energy structure:}
    As the structure is fit in position and amplitude, one obtains uncertainties on both parameters. Both parameters are varied withing their 1$\sigma$ uncertainties.
    
    \item \textbf{Rate decrease:}
    The decrease of the rate through tritium loss over the acquisition time is approximately linear. Thus, the measured rates of the 15~keV monitor points are fit linearly and a correction factor is calculated for each measured point. The linear function parameters are varied within their correlated uncertainties.
    
    \item \textbf{Trapping:}
    The size of the trapping effect depends on the ratio of elastic to inelastic cross section. While the elastic cross-section is assumed to be precise, an uncertainty of 2.5~\% is considered on the inelastic cross-section~\cite{Liu1987, Aseev2000}. This is used as an input parameter to the trapping simulation to propagate the uncertainty into a covariance matrix.
\end{itemize}
The influence of other systematic effects, e.g.~fluctuations in the electrode high voltage or the source column density, is negligible for the given statistical sensitivity of the measurement.

\subsubsection{Fit to the integral spectrum}

\begin{figure}
    \centering
    \includegraphics[width=0.6\textwidth]{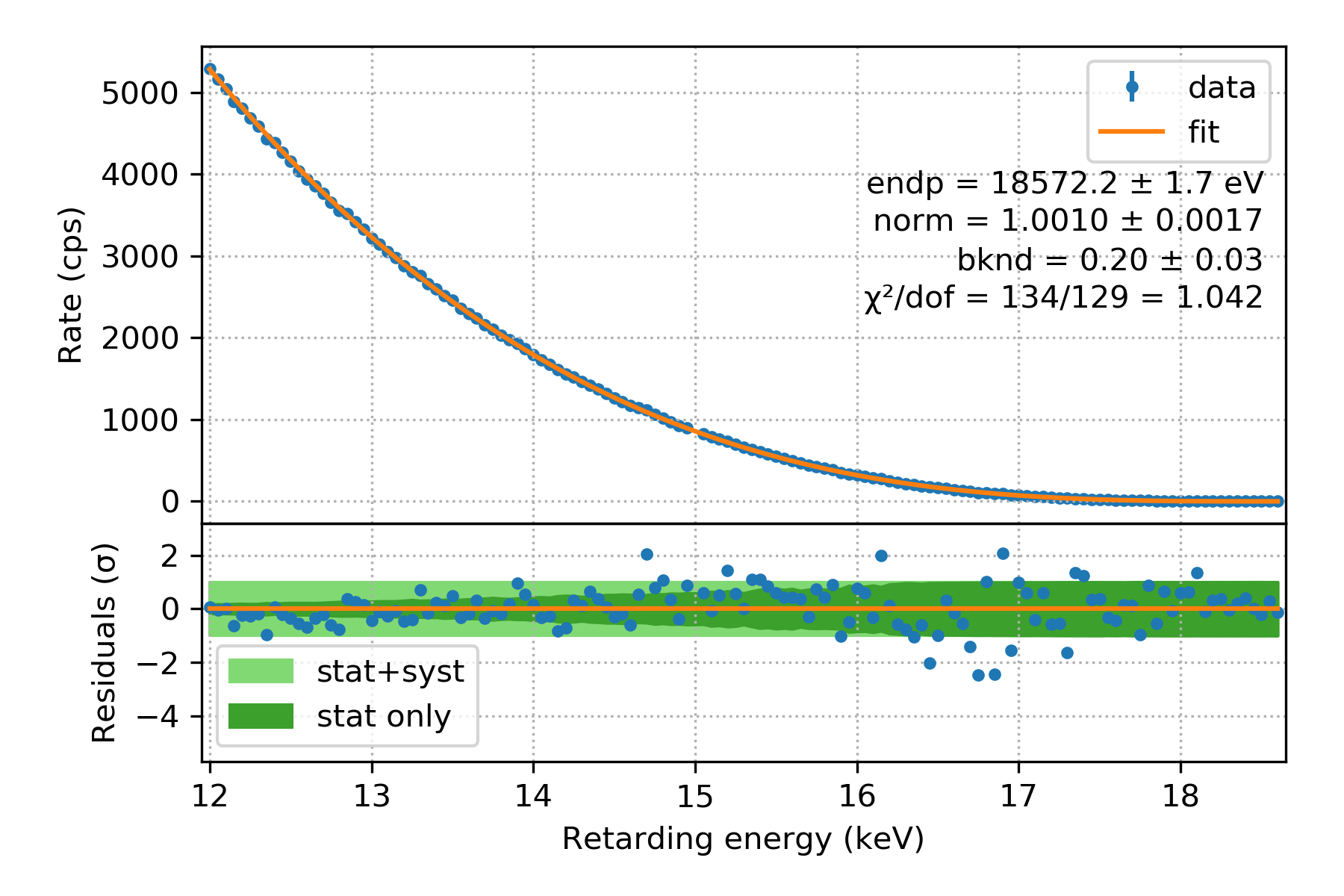}
    \caption{\textbf{Fit of the data taken in integral mode.} The 15~keV monitor points are excluded from the fit. The residuals are normalized to the overall uncertainty that consist of a statistical (dark green) and systematic contribution.}
    \label{fig:fit_int}
\end{figure}

The detector response $D$ does not affect the integral model, since the differential spectrum is integrated over the whole range of possible energies. An average source density over the time of the scan is assumed in the simulation of the response of source and spectrometer $R$. The whole model is given as
\begin{equation}
    f_\mathrm{int.model}(qU; N, E_0, B) = N\cdot\int_{qU}^{E_0}\left[\dv{\Gamma(E,E_0)}{E}\times\Bigl(R(qU,E)\cdot T(E)\Bigr)\right]\mathrm{d}E + B~.
    \label{equ:integral_model}
\end{equation}
Free fit parameters are normalization $N$, spectral endpoint $E_0$, and background $B$. In the final fit a $\upchi^2$ of the form
\begin{equation}
    \upchi^2(N, E_0, B) = \left[y(qU)-f_\mathrm{int.model}(qU; N, E_0, B)\right]^T M^{-1} \left[y(qU)-f_\mathrm{int.model}(qU; N, E_0, B)\right]
    \label{equ:chi2_integral}
\end{equation}
is minimized, where $M$ is a covariance matrix including correlated uncertainties discussed in section~\ref{sec:integral_systematics}. The fit of the model to the data is shown in figure~\ref{fig:fit_int}. Neither endpoint nor normalization nor background show any unexpected behavior. The model is in good agreement with the data with a p-value of 35.5~\%.

\subsection{Analysis of the differential mode}

Only single events in the central pixel are analyzed. The ring of outer pixels is used to veto charge shared events with a coincidence window of 130~ns, as these events distort the measured spectral shape. As the DAQ detection threshold is not a sharp cut, a software threshold of 2~keV was applied.

\subsubsection{Systematic uncertainties}\label{sec:differential_systematics}

The largest systematic uncertainty in the differential measurement mode arises from the detector parameterization. The electrons from the spectrometer electrode have different incident angles to the detector and back-reflection characteristics in the spectrometer than the ones from tritium decay. A deviation in the responses is therefore expected. To take this into account, all response parameters $\phi$ are free in the fit and only constrained by wide pull-terms to guarantee numerical stability during the fit. Other systematic effects on corrections, e.g.~events below the energy threshold or charge sharing, were found to be negligibly small and are not further considered.

\subsubsection{Fit to the differential spectrum}\label{sec:fit_diff}

To build the differential model, the response matrices of trapping $T$, transport $R$ and detector $D$ are convolved with the $\upbeta$-decay model in section~\ref{sec:model}:
\begin{equation}
    f_\mathrm{diff.model}(E; N, E_0, B, \phi) = N\cdot\left[\dv{\Gamma(E,E_0)}{E}\times\Bigl(R(qU,E')\cdot T(E')\cdot D(E',\phi)\Bigr)\right]+B~.
    \label{equ:differential_model}
\end{equation}\\
Free fit parameters are normalization $N$, spectral endpoint $E_0$, background $B$ and the 15 response parameters $\phi$. Furthermore, the parameters $m_E$ and $c_E$ of the energy scale of the model $E'=m_E\cdot E+c_E$ are free to account for inaccuracies in the calibration of the data. In the final fit of the model to the data a $\upchi^2$ of the form
\begin{equation}
    \upchi^2(N, E_0, B, \phi) = \left[y(E)-f_\mathrm{diff.model}(E; N, E_0, B, \phi)\right]^T M^{-1} \left[y(E)-f_\mathrm{diff.model}(E; N, E_0, B, \phi)\right]
    \label{equ:diff_chi2}
\end{equation}
is minimized, where $y$ is the measured tritium spectrum and $M$ is a covariance matrix with only squared statistical uncertainties on the diagonal.\\

\begin{figure}
    \centering
    \includegraphics[width=0.6\textwidth]{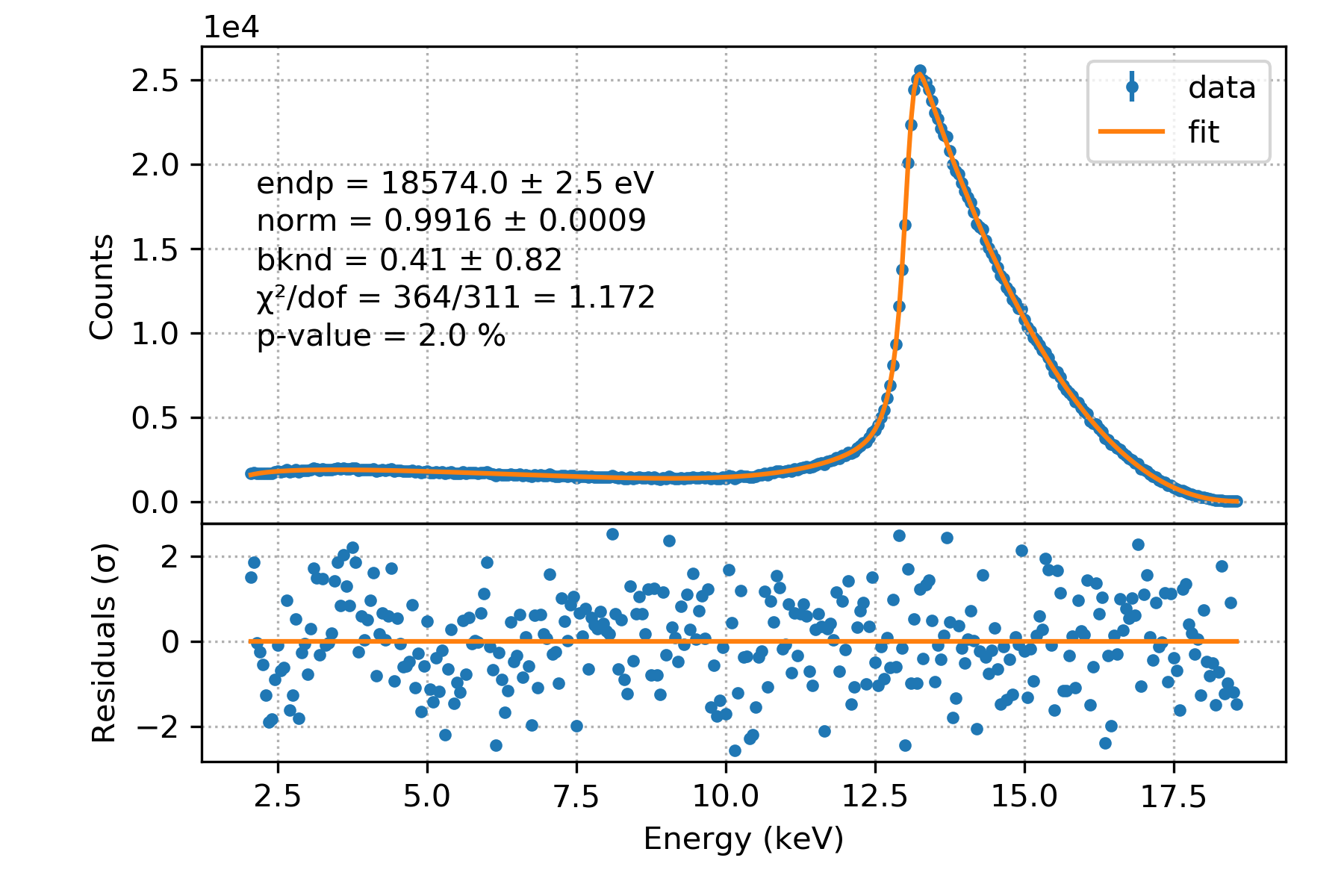}
    \caption{\textbf{Fit of the data taken in differential mode.} The statistical error bars of the data in blue are too small to be seen. The fit is shown in orange. The retarding energy is 13~keV. The residuals are normalized to the statistical uncertainty.}
    \label{fig:fit_diff}
\end{figure}

The fit to the data is shown in figure~\ref{fig:fit_diff}. Considering molecular tritium decay and experimental effects like the work function of the WGTS and the spectrometer, the endpoint is in agreement with the expected value of around 18575~eV~\cite{Otten2008} within uncertainty. Normalization and background show no unexpected behavior. With a p-value of 2.0~\%, the model describes the data in an acceptable way with no apparent residual pattern. Furthermore, the parameters $\phi$ of the response function obtained from the tritium fit reproduce the shape of the spectrometer electrode data. The results of the sterile neutrino search are presented in section~\ref{sec:conclusion}, together with the results of the integral mode.

\subsection{Limit setting}

\begin{figure}
    \centering
    \includegraphics[width=0.6\textwidth]{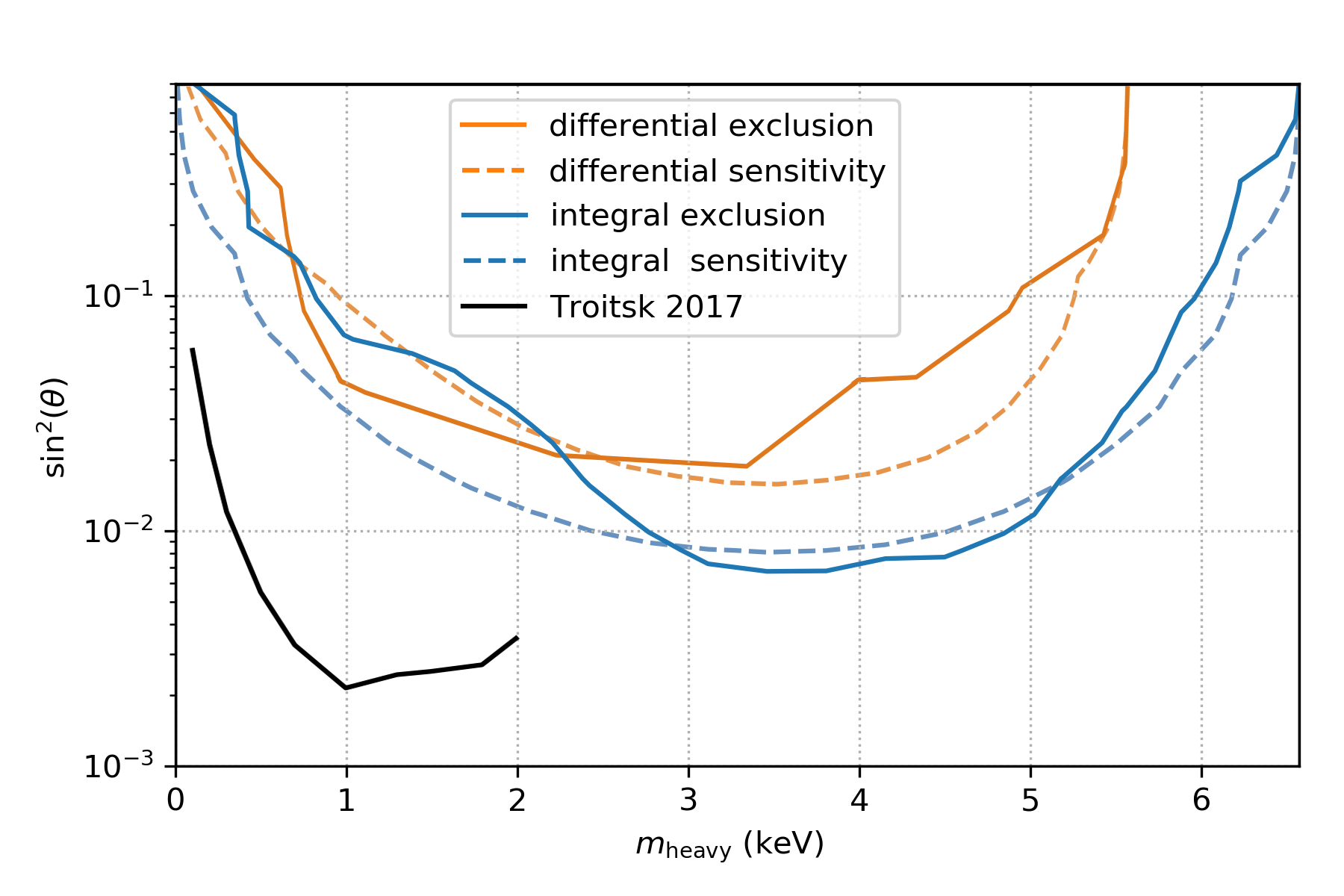}
    \caption{\textbf{Sensitivity and exclusion curves.} For the integral mode, exclusion and sensitivity are shown as straight and dashed lines in blue, respectively. The same holds for the differential mode in orange. The confidence level is 95~\%. The sensitivity provides an expectation for the exclusion based on the given statistical uncertainty. The exclusion is the actual realization given the measured data. The latest result of the Troitsk nu-mass collaboration is shown in black~\cite{Abdurashitov2017}.}
    \label{fig:exclusion}
\end{figure}

To determine the excluded ($m_\mathrm{heavy}$, $\sin^2(\theta)$)-parameter space for a sterile neutrino admixture, an exclusion curve is calculated for both integral and differential mode (see~\cite{Mertens2015} for details). To this end, the fits were performed with a sterile neutrino admixture in the model over a grid of ($m_\mathrm{heavy}$, $\sin^2(\theta)$)-parameters. The 95~\% C.L. exclusion curves are drawn at $\Delta\upchi^2=5.991$ in figure~\ref{fig:exclusion}. The also shown sensitivity curves were derived from fits to so-called unfluctuated MC-data sets~\cite{Cowan2011} and provide an expectation for the exclusion given the level of statistical uncertainty.\\

The explored mass range is determined by the retarding energies that have been measured: down to 6.6~keV in the integral and 5.6~keV below the endpoint in the differential mode. This enlarges the previously analyzed mass range of the Troitsk nu-mass experiment~\cite{Abdurashitov2017} by a factor of 3.\\

The exclusion curves presented are not competitive with other state of the art keV-scale sterile neutrino searches~\cite{Hiddemann1995, Holzschuh1999, Abdurashitov2017}. The purpose of this analysis was to test and compare the analysis of both measurement modes. In both modes, any systematic effect that introduces a kink-like signature that mimics a sterile neutrino admixture is crucial. The integral mode is prone to rate-dependent effects and requires detailed corrections and stable experimental performance. The differential mode is prone to detector effects and requires a complex model of all interactions in the detector and its response. These are largely different systematic effects and none would create a kink in both the integral and differential mode simultaneously. Thus, by comparing the result of both modes, a false-positive signal can be excluded. Neither in the integral nor in the differential exclusion curve, an indication for a sterile neutrino signature is apparent.

\section{Conclusion}\label{sec:conclusion}

In this work a 7-pixel TRISTAN prototype detector was successfully installed at the Troitsk nu-mass experiment. As a technological predecessor of the KATRIN experiment, Troitsk nu-mass offers a very similar experimental environment and hence an ideal test bench for a sterile neutrino search with a TRISTAN detector. The detector performed reliably with an energy resolution around 300~eV (fwhm). Data was taken in two complementary measurement modes, which are prone to different types of systematic effects.\\

A detailed model both for the integral and differential spectrum was developed. It includes a full experimental response function obtained via dedicated calibration measurements and extensive Monte Carlo simulations. Two strategies were developed to analyze the integral and differential spectrum, respectively. In the integral case systematic uncertainties are included with the covariance matrix method, while for the differential case systematic uncertainties are included via nuisance parameters in the $\upchi^2$ fit.\\

A detailed budget of systematic uncertainties was identified and evaluated. As a major result it could be demonstrated that both measurement modes are prone to largely different systematic uncertainties and hence a combination of both allows to exclude possible false-positive signals in future sterile neutrino searches with more data. The analyses in this work exclude the existence of a sterile neutrino with a mixing amplitude of larger than $7\cdot10^{-3}$ (integral) and $2\cdot10^{-2}$ (differential) in the most sensitive mass region, respectively. The accessible mass range of the Troitsk nu-mass experiment was previous enlarged by a factor of 3.\\

The analysis methods developed in this work depict a major milestone in the TRISTAN project and will serve as the basis for future sterile neutrino searches with the TRISTAN detector integrated at the KATRIN experiment.

\acknowledgments{This work was supported by the Max Planck Society (MPG) and the Technical University Munich (TUM). We thank the Halbleiterlabor of the Max Planck Society (HLL), XGLab and the Institute for Nuclear Research of Russian Academy of Sciences (INR) for the fruitful cooperation. Anton Huber and Marc Korzeczek acknowledge the support by the DFG-funded Doctoral School "Karlsruhe School of Elementary and Astroparticle Physics: Science and Technology". The TUM and KIT groups would like to thank Vladislav Pantuev for his generous hospitality during their stays in Troitsk.}


\bibliographystyle{JHEP}
\bibliography{TheBib}

\end{document}